\documentclass[12pt]{article}
\usepackage{amsthm,url}
\usepackage{amsmath}
\usepackage{amssymb}
\usepackage{color}
\usepackage[square,numbers,comma,sort&compress]{natbib}
\usepackage{enumerate}
\usepackage{colortbl}
\usepackage{booktabs}
\usepackage{ifpdf}
\usepackage{graphics}
\usepackage{epsfig}
\usepackage[TABBOTCAP,normalsize]{subfigure}
\usepackage{booktabs}
\usepackage{mathrsfs}  
\usepackage{rotating} 

\usepackage{times}

\usepackage{sectsty}

\sectionfont{\large}
\subsectionfont{\large}

\title{Statistical analysis on detecting recombination sites in DNA-$\beta$ satellites associated with old world geminiviruses}

\begin{document}

\begin{center}
{\large
Statistical analysis on detecting recombination sites in DNA-$\beta$
satellites associated with the old world geminiviruses
}
\end{center}
\vskip 5pt
\begin{center}
Kai Xu\footnote{Department of Plant Pathology, University of Kentucky}
and Ruriko Yoshida\footnote{Ruriko Yoshida,
University Of Kentucky, Department of Statistics, 817 PATTERSON OFFICE TOWER, LEXINGTON KY  40506-0027, ruriko.yoshida@uky.edu}
\end{center}
\begin{abstract}
Although exchange of genetic information by recombination plays an
important role in the evolution of viruses, it is not clear how it
generates diversity. Understanding recombination events helps with
the study of the evolution of new virus strains or new viruses. {\it Geminiviruses} are plant viruses which have
ambisense single-stranded circular DNA genomes and are one of the most
economically important plant viruses in  agricultural production.
Small circular single-stranded DNA satellites, termed DNA-$\beta$,
have recently been found to be associated with some geminivirus infections.
In this paper we analyze several DNA-$\beta$ sequences of
geminiviruses for recombination events using phylogenetic and
statistical analysis and we find that one strain from ToLCMaB has a
recombination pattern and is a recombinant molecule between two
strains from two species, PaLCuB-[IN:Chi:05] (major parent) and
ToLCB-[IN:CP:04] (minor parent). We propose that this recombination
event contributed to the evolution of the strain of ToLCMaB in South
India.  The Hidden Markov Chain (HMM) method developed by Wedd et al estimating phylogenetic tree through out the whole alignment provide us a recombination history of these DNA-$\beta$ strains. It is the first time that this statistic method has been used on DNA-$\beta$ recombination study and give a clear recombination history of DNA-$\beta$ recombination.
\end{abstract}

\section{Introduction}

Geminiviruses are emerging as one of
the most economically important plant viruses in agricultural
production \cite{11, 12, 8_13}. {\it Begomovirus} is the largest genus of the
family of {\it Geminiviridae} and is phylogenetically and geographically
divided into two groups; the Old World viruses and the New World viruses. The new world {\it begomovirus}  consists of two viral genomes,
DNA-A and DNA-B, while most of the Old World begomovirus just has one
partite DNA-A \cite{8_13}. About a decade ago, a satellite molecule called
DNA-$\beta$ was found to associate with some of the old world
geminivirus \cite{14, 15}.

DNA-$\beta$ has a genome approximately 1.3--1.5kb long, and depends on the
helper virus DNA-A for its replication, movement and transmission
\cite{2, 14, 15}. It is grouped into sub-viral agents by the International
Committee on Taxonomy of Viruses (ICTV). The most typical plant
symptoms caused by geminivirus are due to an association of DNA-$\beta$
with DNA-A, whereas DNA-A alone does not lead to severe damage to crops
\cite{2, 8_13}. C1 gene encoded by DNA-$\beta$ were found to suppress host defense systems \cite{36} and modulate host development \cite{37}, and was believed to be one of the determining factors for geminivirus-induced disease symptom development \cite{8_13}.

DNA-$\beta$ has not been found in the New World (North American and
South American continents) and is believed to be associated with Old World
begomoviruses after the geographical divergence of ``Old'' and ``New''
continents \cite{18}. Although DNA-$\beta$ has relatively a large range
of its selection on different species of the helper virus DNA-A \cite{19}, it is proposed to co-evolve with the DNA-A component \cite{8_13}.

Recombination plays an important role in geminivirus \cite{16} and
DNA-$\beta$ evolution \cite{1_17, 20_24}. A fragment of DNA-$\beta$ genome infecting tomato was reported to migrate to cotton via recombination with other adaptive DNA-$\beta$ molecules \cite{1_17}, indicating the role of a recombination event in evolution of DNA-$\beta$ molecules.

Because of the important role of recombination in DNA-$\beta$
evolution, analysis on recombination events of DNA-$\beta$ becomes
specially important for understanding this viral evolution and disease epidemic as well as development of potential control strategies.

In this paper, we apply a statistical phylogenetic analysis using a
Bayesian stochastic method to infer changes in phylogeny along
multiple sequence alignments while accounting for rate heterogeneity
developed by \cite{3} to estimate potential recombination spots of
DNA-$\beta$.  
It is the first time that this statistic method has been used on DNA-beta recombination study and give a clear recombination history of DNA-beta recombination.
In order to confirm our results, we also apply a statistical phylogenetic method developed by \cite{22} to the same data sets.  We find that the results with the method in \cite{3} and with the method in \cite{22} are very similar to each other. One strain of  Tomato leaf curl Maharastra betasatellite (ToLCMaB) has a recombination pattern and is possibly recombinant molecule between two strains from two distinct species, Papaya leaf curl betasatellite (PaLCuB) and Tomato leaf curl betasatellite (ToLCB), PaLCuB-[IN:Chi:05] (major parent) and ToLCB-[IN:CP:04] (minor parent). This recombination event may contribute to the evolution of Tomato leaf curl Maharastra betasatellite.

\section{Data set}

A proposed taxonomy of DNA-$\beta$ using 78\% nucleotide sequence
identity as demarcation threshold was accepted and widely used for
distinguishing species from strains of DNA-$\beta$ \cite{8_13}. This
resulted in about 51 distinct species of DNA-$\beta$ associated with
begomoviruses.

Tomato leaf curl disease (ToLCD) is caused by begomoviruses associated
with betasatellites. A recent report showed that different species of
DNA-$\beta$ associated with ToLCD in India are geographically isolated
and distributed \cite{25}. The DNA-$\beta$ molecules in  southern and central India are more closely related to each other than those in northern India.

To observe potential recombination events among these geographically
related DNA-$\beta$ species, we chose four strains from four distinct
species of DNA-$\beta$ associated with ToLCD in India. Among the four
strains, ToLCBDB-[IN;Luk;05] (taxon-0) and ToLCB-[PK;RYK;97]  (taxon-1)
are from northern India, while PaLCuB-[IN;Chi;05]  (taxon-2) and
ToLCMaB-[IN;Pun;04]  (taxon-3) are from southern India. In the same report
as well as another report \cite{26}, species of ToLCBDB and ToLCB are closely related in phylogenetic tree, while PaLCuB and ToLCMaB are sisters (neighbors).

Another ToLCD associated DNA-$\beta$ from Indonesia  (taxon-4) was
chosen as an out group. Other five species of non-ToLCD related
DNA-$\beta$ from eastern Asia and southeastern Asia (taxa-5, 6, 7,
8, and 9) were also chosen for the out group.  See Table \ref{table1} for details.

\begin{sidewaystable}[!htp]
\begin{center}
\begin{tabular}{l|l|l|l}
taxa & Beta abbreviation/ Accession \# & Full name & Location\\\hline
0 & ToLCBDB-[IN;Luk;05].DQ343289 & Tomato leaf curl Bangladesh
betasatellite & India: Lucknow\\\hline
1 & ToLCB-[PK;RYK;97].AJ316036 &Tomato leaf curl betasatellite &
Pakistan\\\hline
2 & PaLCuB-[IN;Chi;05].DQ118862& Papaya leaf curl betasatellite &
India:Chinthapalli\\\hline
3 & ToLCMaB-[IN;Pun;04].AY838894 & Tomato leaf curl Maharastra
betasatellite & India: Pune, Maharastra\\\hline
4 & ToLCJB-[ID;ID1;03].AB100306  & Tomato leaf curl java betasatellite
& Indonesia\\\hline
5 & AYVB-[CN;Gx96;04].AJ971261 & Ageratum yellow vein betasatellite &
China:Guangxi\\\hline
6 & SibYVB-[CN;Gd13;04].AM230643 & Siegesbeckia yellow vein
betasatellite & China:Guangdong province\\\hline
7 & SibYVGxB-[CN;Gx111;05].AM238695 & Siegesbeckia yellow vein Guangxi
betasatellite & China:Guangxi Province\\\hline
8 & EpYVB-[JR;MNS2;00].AJ438938 & Eupatorium yellow vein betasatellite
& Japan:Fukuoka Prefecture \\\hline
9 & LaYVB-[VN;Hoa;05].DQ641715 & Lindernia anagallis yellow vein
betasatellite & Viet Nam: Hanoi\\
\end{tabular}
\end{center}
\caption{Sequence information of 10 species used in this study}\label{table1}
\end{sidewaystable}

\section{Materials and Methods}

First, a data set of ten DNA-$\beta$ genome sequences in .fasta
format was aligned using clustalw-multialign software with the following
parameters: (Gap opening penalty 10.0, gap extension penalty 0.2, gap
separation penalty range 8, DNA weight matrix: IUB) \cite{5}. 

To analyze recombination for DNA-$\beta$ from geminiviruses, we used the software package from \cite{3}. In this method they applied a hidden Markov model (HMM) to infer changes in phylogeny along multiple sequence alignments while accounting for rate heterogeneity.  Under the HMM, the hidden states are all possible unrooted tree topologies with the number of leaves $n$ fixed along each site.  The observed state space is $\{A, C, G, T, -\}$.  Under the evolutionary model, the evolution of homologous DNA/RNA sequences (or protein-coding sequences where the state space is of size 61) can be described by continuous time Markov chains on a phylogenetic tree. A continuous time Markov chain is characterized by a substitution rate matrix, and the phylogenetic tree summarizes the relationships between the species in terms of edge lengths (times since divergence) and common ancestors. The DNA sequences are only observed in the leaves, and information on the phylogenetic tree, substitution events (time and type) and edge lengths is missing. The transition matrix $P(t)$ for a continuous time Markov process can be written as $\exp(Qt)$, where $Q$ is a parametrized substitution rate matrix which determines the Markov process.
In this method the evolutionary model was set as Hasegawa-Kishino-Yano (HKY) model \cite{4}.    

The rate matrix $Q$ under HKY model is written as the following:
Let $\Sigma = \{A, C, G, T\}$ and let $\pi_a,\; a\in \Sigma, \; \sum_a \pi_a=1,$ denote the stationary distribution of the Markov chain. This distribution can be estimated from the nucleotide frequencies in a single sequence.
HKY model has substitution rate matrix
\begin{eqnarray}
  Q_{\alpha, \beta}= \left[ \begin{array}{cccc}
    \cdot & \alpha\pi_2 & \beta\pi_3 & \beta\pi_4 \\
    \alpha\pi_1 & \cdot & \beta\pi_3 & \beta\pi_4 \\
    \beta\pi_1 & \beta\pi_2 & \cdot & \alpha\pi_4 \\
    \beta\pi_1 & \beta\pi_2 & \alpha\pi_3 & \cdot \\
  \end{array} \right]
  \label{HKY}
\end{eqnarray}
where the diagonal elements are such that each row sums to zero and
the two unknown parameters are $\alpha$ and $\beta$.    The software from \cite{3} estimates the posterior distribution using Monte Carlo Markov Chain (MCMC) method under the HMM and then it outputs each tree topology with its posterior probability along each site (see \cite{3} for details).

We have used HKY model for phylogenetic analysis on our data sets in this paper, since the HMM software in \cite{3} uses HKY model.  Also note that we have used the generalized time reversible (GTR) + gamma + invariant model, which is within the $95\%$ confidence interval computed via Akaike's information criteria (AIC) in the software {\tt jModelTest} \cite{6, 7}, to reconstructing a ML tree and the ML tree under the GTR+gamma+invariant model has the same tree topology as the ML tree under HKY model in Fig. \ref{Fig6} as well as the consensus tree under HKY model in Fig. \ref{Fig5}.

The generated alignment file in phylip format was put in to the HMM
software \cite{3} using the command ``java -jar ST-HMM.jar'' with the following
parameter (iterations: 50000, burn-in: 25000, rates: 0.001, 0.003, 0.01,
0.03, 0.1, 0.3, 1.0, 3.0, 10.0, 100.0, lambda: 5, kappa: 2.0,
tuningpar 0.4). Command ``java -jar STHMMPosterior.jar'' was used to
summarize the posterior distribution, and trees with posterior
probability above 0.05 were selected using the command ``java -jar
TreeSummary.jar''. The region 1--1000 nucleotide (nt) was found to
have a clear pattern of recombination, while the region 1000--1505 nt
seems to have a massive pattern of tree probability.

In order to apply phylogenetic analysis to the sequences of 1--1505 nt
and 1000--1505 nt of the 10 viral sequences after aligning with the
clustalw-multialign software into nexus format, we estimated the
posterior distribution under the generalized time reversible (GTR) +
$\Gamma$ model and HKY model, and we estimated the maximum likelihood
estimators.  First we applied a software {\tt MrBayes} \cite{34}  to
analyze the split of different taxa on the most consensus tree under
the GTR + $\Gamma$ and HKY models.  647300 generations were sampled
for 1-1505 nt alignment, while 3600000 generations were sampled for
1000--1505 nt alignment. The first 25\% of the data was burn-in. We ran
four Markov chains for each model.  We followed the recommendation of
{\tt MrBayes} which suggests running the
chains until the standard deviation of the chains' split frequencies
is less than 0.01.  

In addition, to verify our results we applied the software {\tt RDP3} \cite{22} to the same data sets.  
Sequence alignment in phylip format was used as input for {\tt
  RDP3}. Parameters were set to default used by {\tt RDP3}.   In the software {\tt RDP3} they have implemented several different methods to find recombination sites, {\tt RDP} \cite{21_27}, {\tt GeneConv} \cite{29}, {\tt BootScan} \cite{28}, {\tt MaxChi} \cite{30}, {\tt Chimaera} \cite{31}, {\tt SIScan} \cite{32}, and {\tt 3Seq} \cite{33}.  

The software {\tt RDP} takes basically three steps:  First they
discard non-informative sites from the input data sets and then for
every triplet of taxa, $\{A, B, C\}$,  from the data set, choose the
sister $A$ and $B$.  Second, they use a window of user-defined width
moved among the aligned sub-sequences one nucleotide at a time and
take an average percentage identifying each of the three possible
sequence pairs among $\{A, B, C\}$ at the each position.   Third, the
probability that the nucleotide arrangement in the identified region
that results in $A, B$ appearing more closely related to $C$ may have
occurred by chance is computed using a binomial distribution.

The software {\tt GeneConv} is based on an earlier statistical approach for detecting gene conversion \cite{Sawyer}.  They use the term ‘‘fragment’’ for an aligned or homologous pair of segments in the input alignment. In the process, the highest-scoring fragments in the given alignment are listed and assigned p-values based on the assumption of a random distribution of polymorphic sites. They assign scores as follows: First, all sites that are monomorphic in the alignment are discarded so that only polymorphic sites are considered. Secondly, for a given pair of sequences, matching bases are scored as $+1$ and mismatches as $-m$, where $m$ depends on the pair of sequences. Fragments are assigned p-values similar to the BLAST procedure \cite{Altschul,Karlin}. This p-value is an approximation of the proportion of permutations of the polymorphic sites for which that pair of sequences has some fragment with the observed score or larger \cite{Sawyer}.

The software {\tt BootScan} takes two phases:  ``Scanning phase'' and
``Detection phase''.  In ``Scanning phase'' first they discard
non-informative sites from the input data sets and in each window of
user-defined width move among the given aligned sequences.  It makes
bootstrap samples and compute rooted UPGMAs by definition rooted or mid-pointed neighbor-joining (NJ) trees.  In ``Detection phase'' every combination of triplets is individually examined for bootstrap evidence that one of the sequences may be alternatively more closely related to each of the other two sequences at different positions along its length. The probability that the pattern of sites within a potential recombinant region could have occurred by a chance distribution of mutations is approximated using a Bonferroni corrected version of the binomial distribution.

The software {\tt MaxChi} considers only polymorphic sites: 
For a given position of the moving window on the input sequence
alignment and for a given pair of sequences, a chi-square statistic is
computed to compare two proportions: the proportion of sites at which
the sequences agree in the left half-window and the proportion of
sites at which the sequences agree in the right
half-window. Discordance between these two proportions may reflect a
recombination event in the history of
the two sequences. The maximum chi-square over all sequence pairs is
recorded as a summary of the evidence for recombination at the window
center.  
Significance of observed chi-square statistics is assessed by a Monte Carlo permutation test.

The software {\tt Chimaera} is also a modification of Maynard Smith's maximum $\chi^2$ method \cite{Wiuf} with only variable sites. The statistic is the maximum $\chi^2$ in the original alignment. The p-value equals the number of times the original statistic is smaller than the statistic from permuted alignments divided by the number of permutations. For all calculations, a sliding window was used, with the width of the window set to the number of polymorphic sites divided by 1.5. This window moves in steps of one nucleotide at a time. 

The software {\tt SIScan} uses a similar idea as algorithms
implemented 
in {\tt MaxChi} and {\tt Chimaera}, but instead of using contingency tables they use Gaussian distribution and use Z-score to compute the p-value.

The software {\tt 3Seq} is similar to {\tt RDP}: {\tt 3Seq} discards non-informative sites from the input data sets and then for every triplet of taxa, $\{A, B, C\}$,  from the data set, it chooses the sister $A$ and $B$: two parent sequences that may have recombined, with one or two breakpoints, to form the third sequence (the child sequence). Excess similarity of the child sequence to a candidate recombinant of the parents is a sign of recombination; they take the maximum value of this excess similarity as the test statistic. Then they rapidly calculate the distribution of the excess similarity and using this method they estimate the p-value.

\section{Results}

\begin{figure}[!htp]
\begin{center}
\includegraphics[width=7cm]{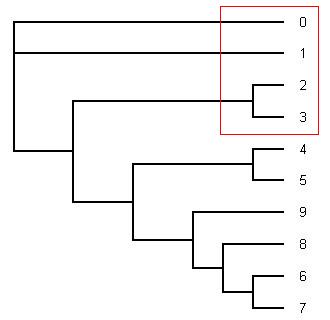}
\end{center}
\caption{This is the tree topology written in pink (series 14) in
  Fig. \ref{Fig1}.  This is an unrooted tree.  This is the most likely
  tree topology from position 1 to 140 and position 300 to
  1000.  The software from \cite{3} and {\tt RDP3} \cite{22} indicate a potential
  recombination event among taxa 0, 1, 2, and 3 in the red rectangle.  Also the ML tree estimated by the software {\tt PHYML} has the same tree topology under HKY model as well as the consensus tree estimated by the software {\tt MrBayes} under HKY and GTR + $\Gamma$.}\label{Fig4} 
\end{figure}

The
most consensus trees found with the 1--1505 nt and 1000--1505 nt
alignment were the same as the most dominant tree found with the HMM
software (the pink tree in Fig. \ref{Fig4}).  

\begin{figure}[!htp]
\begin{center}
\includegraphics[width=12cm]{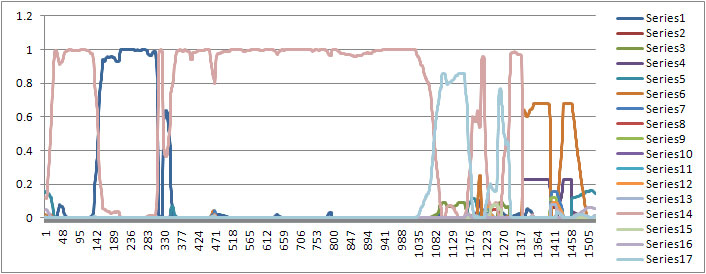}
\end{center}
\caption{The figure shows an estimated probability of each tree
  topology along each site computed using the software from \cite{3}.
  The label of ``Series $i$'' for $i = 1, \cdots , 17$ in the figure
  represents each different tree topology. The y-axis represents the probability for each tree topology and the x-axis represents position number.  The tree written in pink is in Fig. \ref{Fig4} and the tree written in the dark blue dominating from  position 140 to 300 is in Fig. \ref{Fig3}.}\label{Fig1} 
\end{figure}

\begin{figure}[!htp]
\begin{center}
\includegraphics[width=7cm]{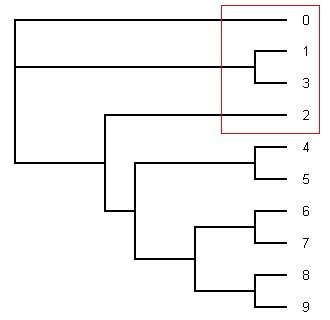}
\end{center}
\caption{The tree written in the dark blue (series 1) in
  Fig. \ref{Fig1}.  This is an unrooted tree. This is the most likely
  tree topology from  position 140 to 300.  The software from
  \cite{3} and {\tt RDP3} \cite{22} indicate a potential
  recombination event among taxa 0, 1, 2, and 3 in the red rectangle.  }\label{Fig3} 
\end{figure}

Then we estimated the maximum likelihood (ML) tree from the whole
alignment (including  position 1 through position 1505).  Next we
infer phylogenetic tree using maximum likelihoods method, using {\tt
  PHYML v3.0} software \cite{7},  with all settings default, namely the evolutionary model is HKY model, the tree topology search operation method is Nearest Neighbor Interchange (NNI), and the starting tree was computed using {\tt BIONJ} \cite{35}.  To analyze the splits of different taxa on the ML tree we applied bootstrapping on the columns of each alignment with the bootstrap sample size 1000.  The ML tree found with the 1-1505 nt alignment was the same as the most dominant tree found with the HMM software (the pink tree in Fig. \ref{Fig4}).

From position 1 to position 141 and from  position 312 to
position 1000, the tree topology in Fig. \ref{Fig4} has almost probability
1.0 (see Fig. \ref{Fig1}).  Note that the estimated ML tree and the estimated
consensus tree reconstructed with the whole sequences from an
estimated posterior distribution have the same tree topology.
However, from  position 141 to position 311 in the alignment,
the tree topology in Fig. \ref{Fig3} has almost probability 1.0 (see Fig. \ref{Fig1}).
The Robinson-Foulds (RF)   distance \cite{10} between the tree
topology in Fig. \ref{Fig3} and tree topology in Fig. \ref{Fig4} is 6.  Note that the
largest possible RF distance for trees with $n$ taxa is $2n - 6$ which
is 14 in our case (the normalized RF distance between these tree
topologies is 0.43).  Thus we do not think this happened because
of the low support of a split but this seems to indicate strongly that
around position 142 and position 311 there are possible
recombination sites.  

\begin{figure}[!htp]
\begin{center}
\includegraphics[width=7cm]{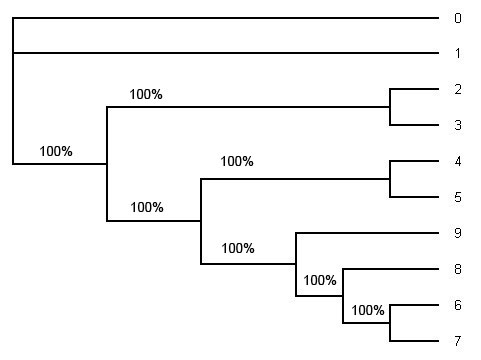}
\end{center}
\caption{The consensus tree estimated by the software {\tt MrBayes} under HKY  from the whole alignment (including position 1 through position 1505).  This is an unrooted tree.  The number in each split represent the probability of the split.  The consensus tree estimated under the GTR + $\Gamma$ also has the same tree topology but it has smaller probabilities of some splits.   Note that the tree topology of the consensus tree is the same as the tree topology of the ML tree in Fig. \ref{Fig6} and  the tree topology in Fig. \ref{Fig4}.}\label{Fig5} 
\end{figure}

\begin{figure}[!htp]
\begin{center}
\includegraphics[width=7cm]{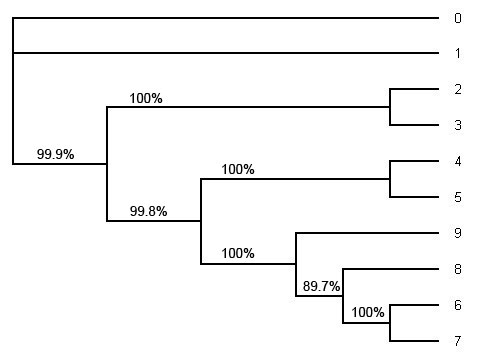}
\end{center}
\caption{The ML tree estimated by the software {\tt PHYML} under HKY model from the whole alignment (including position 1 through position 1505).  This is an unrooted tree. The number in each split represents the probability of the split estimated by bootstrapping with the bootstrap sample size 1000.    Note that the tree topology of the ML tree is the same as the tree topology of the  consensus tree in Fig. \ref{Fig5} and the tree topology in Fig.  \ref{Fig4}.}\label{Fig6} 
\end{figure}

In order to compute the support for each split we have also computed
the consensus tree using the software {\tt MrBayes} (Fig. \ref{Fig5}) and the
ML tree using {\tt PHYML} (Fig. \ref{Fig6}).   For the consensus tree we used
the posterior distribution and for the ML tree we use the bootstrap
with the sample size 1000 to compute the support for each split.  They
have the same tree topology as the tree in Fig. \ref{Fig4} and the support for
each split in the ML tree and the consensus tree has very high
probability.  Especially, the probability of each split on the
consensus tree estimated with the whole sequences under HKY is 1.0
(100\%).  
(
Even though one of the splits on the ML tree reconstructed with the whole sequences under HKY has about 90\% of its support all other splits have strong support (Fig. \ref{Fig6}).)

\begin{figure}[!htp]
\begin{center}
\includegraphics[width=12cm]{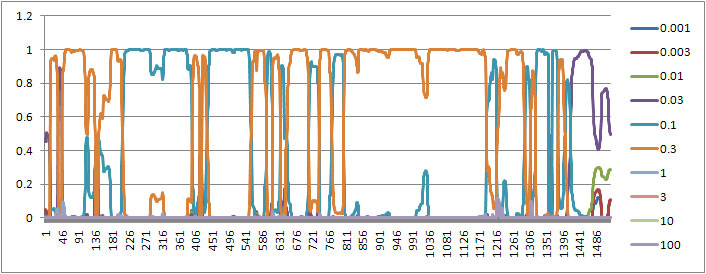}
\end{center}
\caption{The figure shows an estimated probability of each mutation rate along each site computed using the software from \cite{3}.  The y-axis represents the probability for each mutation rate and the x-axis represents position number. It shows that the most common rates are 0.1 and 0.3.}\label{Fig2} 
\end{figure}

 The mutation rates along each site are also estimated by the software
 from \cite{3} and it seems that the mutation rates are between 0.1
 and 0.3 (Fig. \ref{Fig2}). 

\begin{figure}[!htp]
\begin{center}
\includegraphics[width=12cm]{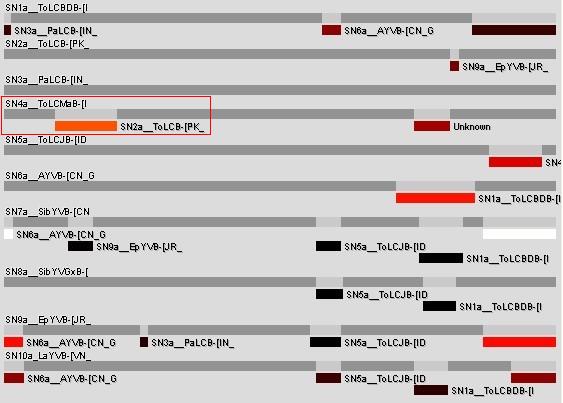}
\end{center}
\caption{{\tt RDP3} infers the same 10 taxa alignment used in our study. Red rectangles indicate the same event inferred by a HMM method from \cite{3}.}\label{Fig7} 
\end{figure}

{\tt RDP3}  estimated a similar recombination event,
where a small genome fragment of ToLCMaB-[IN;Pun;04] (taxon-3)
(position 142-311 in alignment) is migrated from ToLCB-[PK;RYK;97]
(taxon-1), as circled by red rectangle in Fig. \ref{Fig7}.  {\tt RDP3} uses multiple
methods for recombination estimation, and the average p-value from
different methods are listed below (Table \ref{table2}).

\begin{table}[!htp]
\begin{center}
\begin{tabular}{l|l|l}
method & events & av. p-val\\\hline
RDP & 1 & $1.962 \cdot 10^{-13}$\\\hline
GENECONV & 1 & $2.158 \cdot 10^{-9}$\\\hline
BootScan & 1 & $2.073 \cdot 10^{-14}$\\\hline
MaxChi & 1 & $7.397 \cdot 10^{-8}$\\\hline
Chimaera & 1 & $2.830 \cdot 10^{-9}$\\\hline
3Seq & 1 & $4.410 \cdot 10^{-2}$ \\
\end{tabular}
\end{center}
\caption{Average p-value from different methods in {\tt RDP3} \cite{22}
  inferring the recombination event between ToLCMaB-[IN;Pun;04]  and
  ToLCB-[PK;RYK;97]  at position of 142-311.  We used RDP
  \cite{21_27}, GENECONV \cite{29}, BootScan \cite{28},  MaxChi
  \cite{30}, Chimaera \cite{31},  
and 3Seq \cite{33}.  We set parameters for each software as follows;
{\tt RDP}: Reference sequence:no; window size:30; Detect recombination
between sequence identity: 0\%--100\%; {\tt GENECONV}: Sequence
option: Treat blocs as one polymorphism; G-scale:1; Max number of
global frags listed per sequence pair: 2000; Max. number of pairwise
frags listed per sequence pair:0; Min. aligned fragment lenghth :1;
Min polymorphisms in frags:2; Min. pairwise frag score:2; Max. number
overlapping frags:1; {\tt Bootscan}:
window size:200; step size:20; use distances; number of bootstraps
replicates:100; Random number seed:3; cutoff percentage:70;
transversion rate ratio: 0.5; coefficient of variation:1; 
{\tt MaxChi}: Window size:70; Gaps: no; 
{\tt Chimaera}: Window size:60; and  {\tt 3Seq}:
Sequences are circular; Highest acceptable P-value:0.05; Bonferroni
correction; Number of permutations:0; use SEQGEN parametric simulations; }\label{table2}
\end{table}

\section{Conclusion}

We first reported a potential recombination event between taxa 1, 2,
and 3, indicating that the strain ToLCMaB-[IN;Pun;04] (taxon-3) from
ToLCMaB is a recombinant of two strains from two different
species, ToLCB-[PK;RYK;97] (taxon-1) and PaLCuB-[IN;Chi;05]
(taxon-2). As one study reported, ToLCMaB-[IN;Pun;04] (taxon-3) and
PaLCuB-[IN;Chi;05] (taxon-2) are closely related in their phylogeny
compared to other species \cite{25}.  Our study showed that ToLCMaB-[IN;Pun;04]
(taxon-3) shares sequence identity mainly with PaLCuB-[IN;Chi;05]
(taxon-2), while a small portion of its genome (position 141 nt to 312
nt in the alignment)  
is potentially migrated from another species, ToLCB-[PK;RYK;97]
(taxon-1).

\begin{figure}[!htp]
\begin{center}
\includegraphics[width=10cm]{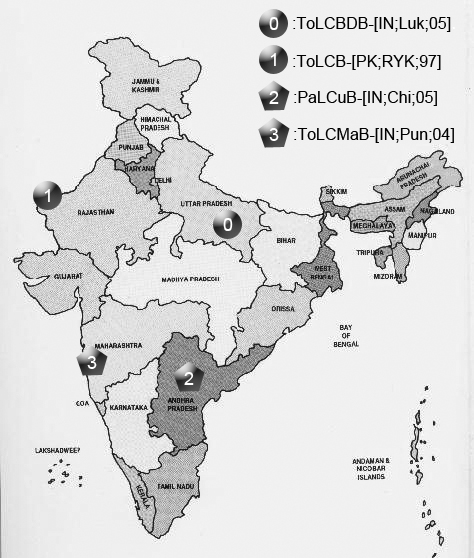}
\end{center}
\caption{The geographic distribution of four betasatellites,
  ToLCBDB-[IN;Luk;05], ToLCB-[PK;RYK;97], PaLCuB-[IN;Chi;05] and
  ToLCMaB-[IN;Pun;04], associated with ToLCD in the India sub-continent.}\label{Fig8} 
\end{figure}

Our results indicate a recombination event happened between a northern
India DNA-$\beta$ strain ToLCB-[PK;RYK;97] (taxon-1)  and a southern
India DNA-$\beta$ strain PaLCuB-[IN;Chi;05] (taxon-2), resulting a new
strain ToLCMaB-[IN;Pun;04] (taxon-3) which was found in southern
India. Different geographic locations provide different physiology of
host, weather conditions, helper viruses, and so on. The 
phylogenetic relationship among ToLCB-[PK;RYK;97] (taxon-1), 
PaLCuB-[IN;Chi;05] (taxon-2), and
ToLCMaB-[IN;Pun;04] (taxon-3) coincides with their distinct geographic
relationship, suggesting that different genetic information on the
viral genome from northern India or southern India may already adapt
to their geographic distribution (Fig. \ref{Fig8}). However, although
the recombination event lead to the possible emergence of a new strain
in a different epidemic location in India, it still has a stronger relationship within its parents geographically and phlegmatically than other strains which are epidemic in other Asian countries.

$\beta$C1 protein, product of the C1 gene, can alter leaf development
and suppress plant defense systems during infection \cite{36, 37}. The
recombination happened in approximate 100-220 nt of the genome (141-312 in alignment), which partially covers the C-Terminal of C1 gene on the beta-satellite. ToLCB-[PK;RYK;97] (taxon-1) $\beta$C1 has 118 amino acids, while $\beta$C1 of PaLCuB-[IN;Chi;05] (taxon-2) has 122 amino acids. The recombination event leads to a $\beta$C1 protein of ToLCMaB-[IN;Pun;04] (taxon-3) with 118 amino acids, missing the 6 amino acids from major partent PaLCuB-[IN;Chi;05] (taxon-2) on the C-terminal of $\beta$C1, instead having 2 amino acids from C-terminal of $\beta$C1 on minor parent ToLCB-[PK;RYK;97] (taxon-1). Although functions of different domains of $\beta$C1 were unknown, the recombination on C-terminal of $\beta$C1 might modulate its function involving in virus-host interaction.

DNA-$\beta$ was known to be capable to adapt to a new helper virus
from distinct geographic location by modifying its genome
\cite{38}. The genetic modification on this southern Indian
DNA-$\beta$ strain ToLCMaB-[IN;Pun;04] (taxon-3) via a recombination
event may contribute to the fitness of 
this DNA-$\beta$ strain on its host.

\section{Discussion}

The advantage of our study is that estimating of phylogenetic tree through out the alignment by HMM method provide a clear history of DNA-beta recombination. It is the first time that researches  on DNA-beta recombination use such statistic method and give this clear recombination history. 

Our study also provides a way to understand DNA virus evolution through
recombination events. From our results, it is likely that the specie
of ToLCMaB is a result of recombination from two different species,
namely ToLCB and PaLCuB. 
Such recombination event contributed to the occurrence of new DNA-$\beta$ species as
well as the evolution of DNA-$\beta$.  
By providing the recombination history together with geographic information, we could link the phylogeny information to the geographic information of DNA-beta strains, thus help us understand evolution and epidemic of the virus.

\section{Acknowledgments}
R. Y. is supported by NIH R01 grant 5R01GM086888.   We thank David
Haws for computations.

\bibliographystyle{elsart-harv}
\bibliography{gemini}

\end{document}